\documentclass[twocolumn,aps,pra,showpacs,floatfix,superscriptaddress]{revtex4-1}
\usepackage{graphicx}
\usepackage{feynmp}
\usepackage{amsmath,amsfonts,amssymb,verbatim,float}

\usepackage[usenames]{color}	
\usepackage{colortbl}

\usepackage{ulem} 

\usepackage{hyperref} 
\usepackage{subfigure}
\usepackage{indentfirst}

\usepackage{makecell}

\usepackage{siunitx} 
\definecolor{faded}{gray}{0.45}

\usepackage{color,xcolor}
\definecolor{BLUE}{rgb}{0.0,0.0,1.0}

\usepackage{ulem}
  
\newcommand{\stA}{\left(2s 2p_{1/2}\right)_0}
\newcommand{\stB}{\left(2p_{1/2} 2p_{3/2}\right)_1}
\newcommand{\stC}{\left(2s 2p_{3/2}\right)_2}

\begin{document}
\thispagestyle{empty}
\title{
Complex-scaled {\it ab initio} QED approach to autoionizing states
}
\author{V.~A.~Zaytsev}
\affiliation{
Department of Physics, St.~Petersburg State University,
Universitetskaya 7/9, 199034 St.~Petersburg, Russia
}
\author{A.~V.~Malyshev}
\affiliation{
Department of Physics, St.~Petersburg State University,
Universitetskaya 7/9, 199034 St.~Petersburg, Russia
}
\author{V.~M.~Shabaev}
\affiliation{
Department of Physics, St.~Petersburg State University,
Universitetskaya 7/9, 199034 St.~Petersburg, Russia
}
\begin{abstract}
\textit{Ab initio} method based on a complex-scaling approach and aimed at a rigorous QED description of autoionizing states is worked out.
The autoionizing-state binding energies are treated nonperturbatively in $\alpha Z$ and include all the many-electron QED contributions up to the second order. 
The higher-order electron correlation, nuclear recoil, and nuclear polarization effects are taken into account as well. 
The developed formalism is demonstrated on the $LL$ resonances in heliumlike argon and uranium. 
The most accurate theoretical predictions for the binding energies are obtained.
\end{abstract}
\maketitle
\begin{figure}[b!]
\begin{center}
\includegraphics[width=0.3\textwidth]{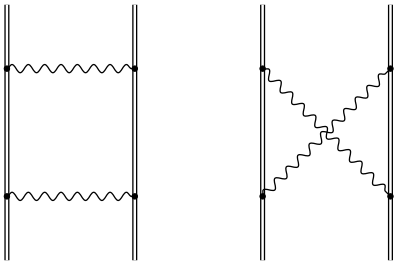}
\caption{
Two-photon exchange diagrams: the ladder (left) and crossed (right) contributions.
The double line represents the electron propagator in the nucleus field and the wavy line designates the photon propagator.
}
\label{feynman_diags}
\end{center}
\end{figure}
\begin{figure*}
\begin{center}
\includegraphics[width=0.9\linewidth]{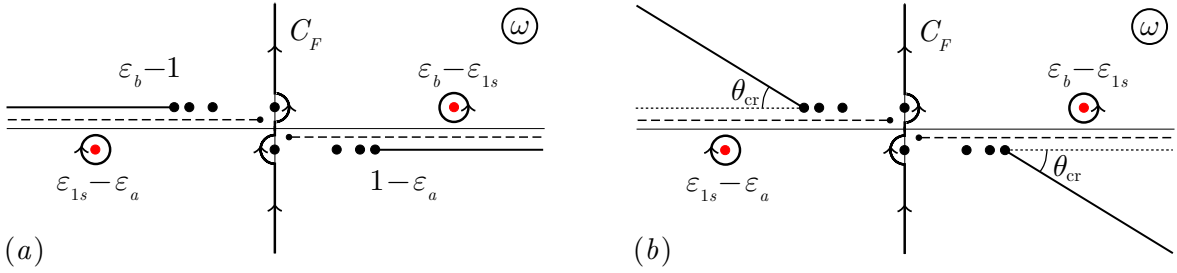}
\caption{
Poles and branch cuts of the integrand and the integration contour for the direct part of the irreducible ladder contribution for $\varepsilon_{1s} < \varepsilon_a \leqslant \varepsilon_b$. 
The solid line representing the continuum turns into a set of poles for the finite basis.
The dashed lines denote the cuts due to the photon propagator.
}
\label{fig_pole_structure}
\end{center}
\end{figure*}
\begin{figure}
\begin{center}
\includegraphics[width=0.5\textwidth]{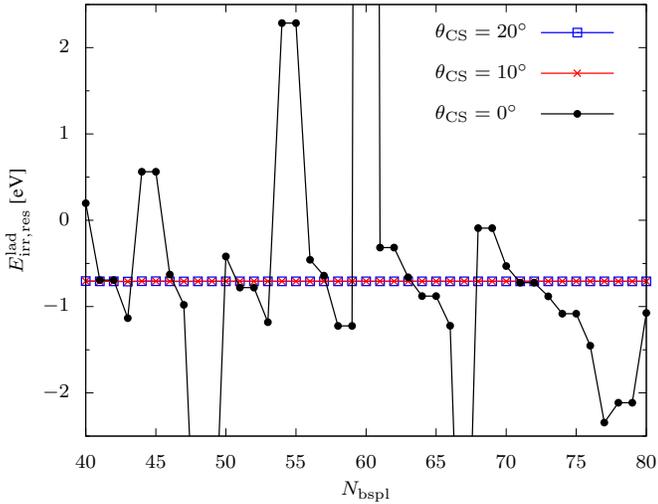}
\caption{
The total resonant contribution, $\left(\kappa_{n_1},\kappa_{n_2}\right)=(-1,1)$ and $\left(\kappa_{n_1},\kappa_{n_2}\right)=(1,-1)$, for the state $\stA$ in uranium as a function of $N_{\rm bspl}$ being the number of the $B$ splines.
}
\label{fig_dependence}
\end{center}
\end{figure}

%
%
Autoionizing states of atomic or ionic systems are the excited states which can decay by virtue of the electron-electron interaction via the emission of one or more electrons. 
The high-precision energies of such states are in demand for plasma diagnostics~\cite{Beiersdorfer_1993, Widmann_1995, kunze_2009, tallents_2018} e.g., in fusion facilities~\cite{Fabian_1991} and astrophysical searches~\cite{Porquet2010}. 
Furthermore, the possibility of using the autoionizing states as the energy-reference standards at synchrotron-radiation facilities is currently investigated~\cite{Simon_PRL105_183001, M_ller_2015, Schippers_2016, M_ller_2018}.
\\
\indent
To obtain the energies of the autoionizing states with the precision necessary for all these and many other applications, accurate calculations of the electron-electron correlations and quantum-electrodynamics (QED) corrections are required. 
State-of-the-art QED calculations constitute an extremely difficult task, and previously they were performed only for the ground and singly-excited states, see Refs.~\cite{Sapirstein:2008:25, Shabaev:2008:1175, Glazov:2011:71, Volotka:2013:636, Shabaev:2018:60, Indelicato:2019:232001} for review.
These calculations usually employ finite-basis-set approaches which fail being naively applied to the computation of autoionizing-state energies.
The failure originates from the fact that the autoionizing states are embedded into the positive-energy continuum, which is discretized in the finite-basis-set approaches.
Discretization leads to the problem of small denominators in certain many-electron QED corrections as well as in correlation contributions treated perturbatively.
As a result, the convergence of the correlation and QED corrections with respect to the size of the basis set is weak or even absent that strongly limits the accuracy.
Moreover, the bases are often constructed from the square-integrable functions that do not properly describe the non-localized autoionizing states.
This, in turn, limits the accuracy of the nonperturbative many-electron methods such as, e.g., configuration-interaction~(CI) and coupled-cluster~(CC) ones.
\\
\indent
The problems associated with the embedding of the autoionizing states into the continuum can be naturally solved by the complex-scaling~(CS) approach, in which the Hamiltonian is dilated into the complex plane.
The autoionizing states corresponding to the dilated Hamiltonian detach from the continuum and admit description by square-integrable functions.
As a result, the CS provides an opportunity to utilize the standard well-established techniques with minor modifications. 
A detailed description of the CS approach as well as its various applications can be found in Refs.~\cite{HO_1983, MOISEYEV_1998, Moiseyev_2011, LINDROTH_2012}.
\\
\indent
The CS approach combined with the perturbation theory~\cite{Lindroth_1994_49, Lindroth_1995_52, M_ller_2018}, Hylleraas~\cite{Ho_1981_23}, Hylleraas-CI~\cite{Pestka_2006_39, Pestka_2007_40, Bylicki_2008_77}, CI~\cite{Brandefelt_1999_59, Derevianko_2010_82, Zhang_2012_85, Zhang_2012_1023, Geng_2016_144, Zaytsev_PRA100_052504_2019, Zaytsev_2020_128}, CC~\cite{Sajeev_2014_33, Matz_2022_156}, and multiconfigurational self-consistent field~\cite{Yeager_2005_104, Samanta_2008_112, Samanta_2012} methods has been successfully applied to the evaluation of the autoionizing-state energies.
In all these calculations, the QED corrections were at best only estimated or not even taken into account that strongly limited the accuracy of the results.
As far as we know, {\it ab initio} QED description of the states being in resonance with the continuum has not been undertaken until now.
Here we combine the rigorous QED treatment with the CS approach and calculate the complete set of many-electron QED corrections to the energies of autoionizing states.
We also account for the one- and two-loop QED contributions, nuclear recoil effect, and higher-order correlation and QED corrections.
The developed approach is applied to the lowest nonmixing autoionizing states of several heliumlike ions, namely, to the $\left(2s 2p_{1/2}\right)_0$, $\left(2p_{1/2} 2p_{3/2}\right)_1$, and $\left(2s 2p_{3/2}\right)_2$ levels in ${\rm Ar}^{16+}$ and ${\rm U}^{90+}$.
\\
\indent
We use the Dirac equation as a zeroth-order approximation and utilize the Furry picture, in which the electron-nucleus interaction is treated nonperturbatively.
The electron-electron correlation and QED contributions are accounted for by perturbation series.
We consider all the contributions to the binding energies up to the second order that to date correspond to the most advanced bound-state QED calculations. 
Below we present the details of the method.
\\
\indent
Let us start with the two-photon exchange contribution, whose evaluation causes the main difficulties for autoionizing states (the calculations for nonresonant states were performed, e.g., in Refs.~\cite{Blundell:1993:2615, Lindgren:1995:1167, Mohr:2000:052501, Yerokhin:2001:032109, Andreev:2001:042513, Asen:2002:032516, Andreev:2003:012503, Andreev:2004:062505, Artemyev:2005:062104, Kozhedub:2019:062506, Yerokhin:2022:022815}).
For two-electron system, this correction is given by the Feynman diagrams shown in Fig.~\ref{feynman_diags}.
\noindent
The ladder contribution is naturally divided into the irreducible and reducible parts~\cite{Shabaev:1994:4489, Shabaev_PR}.
For nonmixing states, the irreducible part reads as (relativistic units, $\hbar = c = m_e = 1$, where $m_e$ is the electron mass, are employed)
\begin{widetext}
\begin{equation}
E^{\rm lad}_{\rm irr} =
\sum_{P} (-1)^{P} {\sum_{n_1 n_2}}' \, \frac{i}{2\pi} \int \! d\omega \, 
\frac{
\left\langle Pa Pb \left\vert I\left(\Delta - \omega\right) \right\vert n_1 n_2 \right\rangle
\left\langle n_1 n_2 \left\vert I\left(\omega\right) \right\vert a b \right\rangle
}
{
\left[\varepsilon_a + \omega - \varepsilon_{n_1}(1 - i0)\right]
\left[\varepsilon_b - \omega - \varepsilon_{n_2}(1 - i0)\right]
} \, ,
\label{eq_irr_lad}
\end{equation}
\end{widetext}
where $P$ is the permutation operator, $(-1)^P$ is its parity, $I$ is the interelectronic-interaction operator, $\varepsilon_n$ is the Dirac energy of the one-electron orbital~$\left\vert n \right\rangle$, and $\Delta = \varepsilon_{Pa} - \varepsilon_a$. 
The summation over $n_1$ and $n_2$ is restricted by the condition $\varepsilon_{n_1} + \varepsilon_{n_2} \neq \varepsilon_a + \varepsilon_b$.
For autoionizing states, $\varepsilon_a + \varepsilon_b > \varepsilon_{1s} + 1$, and this condition is fulfilled when one of the $n_1$ and $n_2$ states equals $1s$ and the other lies in the continuum.
Nevertheless, such the intermediate states should be attributed to the summation in Eq.~\eqref{eq_irr_lad}.
For the sake of simplicity, Eq.~\eqref{eq_irr_lad} is given for a one-determinant state.
The transition to the general case of a many-determinant wave function is straightforward and does not pose additional issues.
\\
\indent
Let us discuss in details the computational difficulties occurring for the irreducible ladder contribution.
In Fig.~\ref{fig_pole_structure}(a), we present the pole structure for its direct part given by the term $(PaPb) = (ab)$ in Eq.~\eqref{eq_irr_lad}.
Upon the Wick's rotation, convenient from the practical point of view, the poles of the electron Green's function are picked up as residues.
For autoionizing states, the residues embedded into the continuum arise, see Fig.~\ref{fig_pole_structure}(a), that causes the failure of the conventional finite-basis-set techniques.
Indeed, in these approaches, the continuum is discretized by the quasicontinuum states with energies depending on the parameters of the basis set.
As a result, the energy difference $\varepsilon_a + \varepsilon_b - \varepsilon_{1s} -\varepsilon_n$ appearing, e.g., in the denominator of the residue for the $LL$ resonances may become arbitrarily small, thus leading to large numerical instabilities.
The same difficulty takes place for the exchange part as well.
We note that the issue of arbitrary small denominators can be avoided in the approaches based on the exact Dirac Green's function.
These approaches, however, are much more complicated in implementation and are not usually utilized for the evaluation of the two-photon exchange diagrams.
\\
\indent
We apply the CS approach to detach the poles from the continuum and overcome the stated difficulties.
We utilize the simplest variant of the CS, namely, the uniform one, in which the radial variable $r$ is transformed according to
\begin{equation}
r \rightarrow re^{i\theta_{\rm CS}} \, ,
\end{equation}
where $\theta_{\rm CS}$ is the CS angle.
This transformation allows us to construct the analytic continuation of the Dirac Hamiltonian into the complex plane.
The discrete spectrum of the dilated Hamiltonian does not change, whereas the continuum spectrum ``rotates''. 
As a result, the pole structure of the integrand in Eq.~\eqref{eq_irr_lad} changes, as schematically depicted in Fig.~\ref{fig_pole_structure}(b).
The CS separates the problematic poles from the continuum and, therefore, eliminates the issue of small denominators making the calculations numerically stable.
\\
\indent
In fact, the CS is required only for terms in Eq.~\eqref{eq_irr_lad} with Dirac quantum numbers $\kappa_{n_1}$ and $\kappa_{n_2}$ that can form an intermediate state into which the Auger decay is allowed.
For instance, the Auger-decay channel of the $\stA$ state is $\stA \rightarrow 1s + \varepsilon p_{1/2}$ that corresponds to $\left(\kappa_{n_1},\kappa_{n_2}\right)=(-1,1)$ and $\left(\kappa_{n_1},\kappa_{n_2}\right)=(1,-1)$.
Hereafter, we refer to such terms as the resonant ones.
When applying the CS, one needs to take the CS angle from the range $\theta_{\rm cr} < \theta_{\rm CS} < \pi/2$, where $\theta_{\rm cr}$ is the critical angle defined by the parameters of the Auger-decay channel~\cite{Balslev1971, Simon1972, Simon1973}. 
For the complete basis set, the value of $\theta_{\rm CS}$ within this interval does not affect the computation results.
In practice, however, incomplete basis sets are used, and one has to find an optimal $\theta_{\rm CS}$ that minimizes the result-variation rate~\cite{Zaytsev_PRA100_052504_2019, Zaytsev_2020_128}.
\\
\indent
We use the finite basis with the basis functions constructed from the $B$ splines~\cite{Sapirstein_1996, Bachau_2001} within the dual-kinetic-balance approach~\cite{Shabaev_PRL93}.
To investigate the dependence of the irreducible ladder contribution on the parameters of the basis set with and without the CS, in Fig.~\ref{fig_dependence} we present the resonant terms for the $\stA$ state as a function of the $B$-spline number~$N_{\rm bspl}$ for various values of $\theta_{\rm CS}$.
\noindent
It is seen that the direct approach with $\theta_{\rm CS}=0^{\circ}$ fails.
The CS approach, meanwhile, allows us to obtain the results which are free from the irregular dependence on $N_{\rm bspl}$.
We note also that inaccuracy in the choice of $\theta_{\rm CS}$ can be easily compensated by extending the basis set and does not influence strongly the accuracy of the results.
Finally, we point out that the reducible part of the ladder contribution, the crossed diagram, and other second-order many-electron QED corrections do not contain additional difficulties for autoionizing states and can be evaluated directly.
The relevant formulas can be found in, e.g., Refs.~\cite{Shabaev:1994:4489, Yerokhin:1999:3522, Artemyev:1999:45, Kozhedub:2010:042513, Malyshev2017}. 
The developed formalism can be readily adapted for ions with more electrons as well as for the mixing states.
\\
\indent
In Table~\ref{tab_tot}, we present the energies of the autoionizing $\stA$, $\stB$, and $\stC$ states in heliumlike argon and uranium.
%
\begin{table*}
\centering
\caption{
Individual contributions to the energies of the autoionizing $\stA$, $\stB$, and $\stC$ states in heliumlike argon ($Z = 18$) and uranium ($Z=92$) ions.
See the text for the details.
}
\resizebox{\linewidth}{!}{%
\begin{tabular}{
l|   
S[table-format=-4.6(2), scientific-notation=fixed, fixed-exponent=0, round-mode=places,round-precision=6]   
S[table-format=-4.6(2), scientific-notation=fixed, fixed-exponent=0, round-mode=places,round-precision=6]
S[table-format=-4.6(2), scientific-notation=fixed, fixed-exponent=0, round-mode=places,round-precision=6]
|
S[table-format=-5.3(2), scientific-notation=fixed, fixed-exponent=0, round-mode=places,round-precision=3]   
S[table-format=-5.3(2), scientific-notation=fixed, fixed-exponent=0, round-mode=places,round-precision=3]
S[table-format=-5.3(2), scientific-notation=fixed, fixed-exponent=0, round-mode=places,round-precision=3]
}
{} & \multicolumn{3}{c}{$Z = 18$} 
   & \multicolumn{3}{|c}{$Z = 92$}  \\
{} & { $\left(2s 2p_{1/2}\right)_0$ } 
   & { $\left(2p_{1/2} 2p_{3/2}\right)_1$ } 
   & { $\left(2s 2p_{3/2}\right)_2$ }
   & { $\left(2s 2p_{1/2}\right)_0$ } 
   & { $\left(2p_{1/2} 2p_{3/2}\right)_1$ } 
   & { $\left(2s 2p_{3/2}\right)_2$ }
\\
\hline
Dirac    
& -2216.113946 & -2211.309654 & -2211.308511 & -68388.842 & -63860.911 & -63827.605 \\
1ph
&    65.614133 &    80.880215 &    65.354183 &    440.134 &    492.711 &    379.773 \\ 
1 loop   
&     0.148270 &     0.001855 &     0.158836 &     56.630 &     15.639 &     58.568 \\
2ph (Br.)   
&    -0.766652 &    -1.095086 &    -0.762486 &     -1.318 &     -1.853 &     -0.974 \\
2ph (QED)   
&     0.000033 &     0.000064 &    -0.000006 &      0.087 &      0.051 &      0.020 \\
ScrQED   
&    -0.003329 &    -0.000044 &    -0.003828 &     -0.442 &     -0.215 &     -0.343 \\
2 loop   
&    -0.000086 &     0.000009 &    -0.000103 &     -0.248 &     -0.018 &     -0.258 \\
$\geqslant$3ph (Br.) 
&     0.005489 &     0.003367 &     0.005577 &     -0.000 &      0.006 &      0.002 \\
Recoil (Br.) 
&     0.029503 &     0.029469 &     0.029440 &      0.146 &      0.141 &      0.135 \\
Recoil (QED) 
&     0.000044 &    -0.000004 &     0.000044 &      0.058 &      0.012 &      0.050 \\
Nuc. Pol.   
&    -0.000025 &     0.000000 &    -0.000025 &     -0.044 &     -0.004 &     -0.039 \\
ScrScrQED   
&    -0.000017 &    -0.000013 &    -0.000001 &      0.012 &      0.002 &      0.000 \\
\hline
Total                                   
& -2151.086583 (82) & -2131.489821 (16) & -2146.526880 (97) & -67893.833 (97)   & -63354.439 (19)   & -63390.654 (95) \\
Ref.~\cite{Zaytsev_PRA100_052504_2019}  
& -2151.087    (15) & -2131.4900   (19) & -2146.527    (15) &                   &                   & \\ 
Ref.~\cite{Andreev2009}                 
&                   &                   &                   & -67892.71  (30)   & -63353.44  (25)   & -63389.68  (30) \\
\end{tabular}%
}
\label{tab_tot}

\end{table*}

%
The nuclear-charge distribution is described by the Fermi model. 
The values of the fundamental constants are taken from Ref.~\cite{codata}.
The interelectronic-interaction and QED corrections are treated in the Furry picture.
To zeroth order, the energy equals the sum of the one-electron Dirac energies.
The contribution of the first order is provided by the one-photon exchange and one-loop QED diagrams.
We also account for the complete set of the second-order corrections.
The two-photon exchange contribution is evaluated within the {\it ab initio} QED approach combined with the CS as described above.
We divide it into the Breit and QED contributions, see, e.g., Ref.~\cite{Malyshev:2014:062517}.
The two-electron self-energy and vacuum-polarization corrections (the row ``ScrQED'') are calculated using the approaches described in Ref.~\cite{Malyshev2017}.
The one- and two-loop one-electron QED contributions are taken from Ref.~\cite{Yerokhin_JPCRD}.
The contribution arising from the exchange by three and more photons are accounted for within the Breit approximation.
To this end, we utilize the CS version of the Dirac-Coulomb-Breit (CS-DCB) Hamiltonian and calculate the total energies by means of the CI approach, see Ref.~\cite{Zaytsev_PRA100_052504_2019}.
The desired higher-order contribution is extracted following the procedure from Refs.~\cite{Kozhedub2008, Artemyev2013, Malyshev2017}.
The nuclear recoil contribution is also divided into the Breit and QED parts.
Within the Breit approximation, it is treated by employing the mass-shift operator~\cite{Shabaev1985, Shabaev:1988:69, Palmer:1987:5987, Shabaev1998} included into the CS-DCB Hamiltonian. 
We dilate the mass-shift operator into the complex plane, which differs the present approach from the one utilized in Ref.~\cite{Zaytsev_PRA100_052504_2019}, where the conventional (Hermitian) operator was used.
This allows us to obtain considerably more stable numerical results.
The QED nuclear recoil contribution is calculated to zeroth order in $1/Z$, see, e.g., Refs.~\cite{Shabaev1985, Shabaev:1988:69, Shabaev1998, Pachucki:1995:1854, Artemyev:1995:1884, Artemyev:1995:5201, Adkins:2007:042508, Malyshev:2018:085001}.
Finally, we account for the nuclear polarization~\cite{Plunien:1991:5853, Plunien:1995:1119:1996:4614:join_pr, Nefiodov1996, Yerokhin_JPCRD} and deformation~\cite{Kozhedub2008} effects.
To estimate the higher-order QED corrections (the row ``ScrScrQED''), we use the model-QED operator~\cite{Shabaev2013} realized in the \texttt{QEDMOD} package~\cite{Shabaev:2015:175:2018:69:join_pr} and follow the procedure from Ref.~\cite{Malyshev_prl126_183001_2021}.
For argon, the dominant uncertainties arise from the uncalculated higher-order QED corrections.
In the case of uranium, the largest uncertainty comes from the nuclear size effect and two-loop QED correction~\cite{Yerokhin_JPCRD}.
\\
\indent
In Table~\ref{tab_tot}, we also compare our results with the previous calculations.
For argon, an excellent agreement is observed with the previous most-accurate results~\cite{Zaytsev_PRA100_052504_2019} but the present ones are much more precise.
For uranium, in contrast, the obtained energies differ from the ones given in Ref.~\cite{Andreev2009} by 3 to 4 standard deviations depending on the state.
In Ref.~\cite{Andreev2009}, the conventional CI approach with the basis constructed from the $B$ splines and supplemented with the continuum wave functions was applied. 
The uncertainties in Ref.~\cite{Andreev2009} were determined from the dependence of the energies on the basis-set parameters.
As was shown in Ref.~\cite{Zaytsev_PRA100_052504_2019}, this approach can converge to an incorrect value due to improper accounting of the interaction with the continuum, which, in turn, leads to a wrong estimation of the uncertainty. 
\\
\indent
To summarize, we worked out an effective and reliable method based on the complex-scaling approach and aimed at describing the autoionizing states within the {\it ab initio} QED formalism. 
To demonstrate the developed approach, we evaluated the energies of the autoionizing $\stA$, $\stB$, and $\stC$ states in heliumlike argon and uranium rigorously accounting for all the QED contributions up to the second order of the perturbation theory.
%
%
The obtained energies are several orders of magnitude more precise than the previous most-accurate values.
Theoretical predictions of this level of accuracy being supplemented with the experimental data provide a new kind of opportunities to test bound-state QED effects.
Moreover, we believe that the combination of the complex scaling with the rigorous QED theory can be utilized for the evaluation of the QED effects in the external and even supercritical electromagnetic fields.
\\
\indent
This study was supported by the grant of the Russian Science Foundation No. 22-22-00370.
%
%

%
\end{document}